     	\documentclass[twocolumn,aps]{revtex4}

\usepackage{graphicx}
\usepackage{dcolumn}
\usepackage{bm}

\begin{document}

\title{SmB$_6$: A material with anomalous energy distribution function of charge carriers?  }

\author{I. Batko}   
  \affiliation{Institute  of  Experimental  Physics,
 Slovak   Academy  of Sciences, Watsonova 47,
 040~01~Ko\v {s}ice, Slovakia}           

\author{M. Batkova}%
 \email{batkova@saske.sk}  
 \affiliation{Institute  of  Experimental  Physics,
 Slovak   Academy  of Sciences, Watsonova 47,
 040~01~Ko\v {s}ice, Slovakia}

\date{\today}

\begin{abstract} 

		We argue that because of valence-fluctuation caused dynamical changes (fluctuations) of impurity energies in an impurity band of valence fluctuating semiconductors 
both occupied and unoccupied sites can be found in the impurity band 
above as well as below the Fermi level even in the ground state.
		As  a consequence, the ground state energy distribution function  
		of the  subsystem of localized charge carriers for valence fluctuating semiconductors 
		is qualitatively 	different than one for conventional semiconductors at $T=0$~K, 
what sheds new light on interpretation of experimental results of valence fluctuating semiconductors, 
e.g. SmB$_6$ and YbB$_{12}$, at lowest temperatures.
\end{abstract}%
\pacs{}

\maketitle   

SmB$_6$  is a prototypical mixed valence material revealing properties of a narrow-gap semiconductor 
	down to few Kelvins  
\cite{Wachter93,Riseborough00,Allen79,Cooley95,Batko93,Gorshunov99,Batkova18PhB,Stern17,Batkova18Jal,Batkova06}.
Paradoxically, at lowest temperatures its conductivity     
exhibits presence of a temperature non-activated term
\cite{Wachter93,Riseborough00,Allen79,Cooley95,Batko93,Gorshunov99,Batkova18PhB,Stern17,Batkova18Jal}, 
which however, cannot be attributed 
to any scattering scenario known for metals, 
because  very high value of corresponding residual resistivity  
would require superunitary scattering  \cite{Wachter93,Riseborough00,Allen79,Cooley95}. 
According to Mott-Ioffe-Regel viewpoint, conventional Boltzmann transport theory 
becomes meaningless if the characteristic mean free path 
of the itinerant conduction electrons becomes comparable to, 
or less than the interatomic spacing \cite{Ioffe60,Mott71}, 
so the requirement for superunitary scattering implicates that  		
		either (i) electrical conductivity is not homogeneous in the volume, 
		i.e. material contains metallic regions forming  along the sample a conductive path 
			responsible for electrical conductivity at lowest temperatures 
		\cite{Dzero10,Dzero12,Lu13,Alexandrov13,Zhu13} 
or (ii) electrical transport at lowest temperatures is realized via a hopping-type transport 
that in contrast to one in conventional semiconductors 
has to be temperature non-activated \cite{BaBa14}. 

The first mentioned approach supposes  metallic surface of SmB$_6$ 
	 which can be either of topological nature \cite{Dzero10,Dzero12,Lu13,Alexandrov13}
or due to "trivial" polarity-driven surface states \cite{Zhu13}. 
Indeed,
 many experimental observations indicate metallic surface transport in (stoichiometric) SmB$_6$ 
\cite{Wolgast13,Kim13,Syers15}.
On the other hand,
Raman scattering studies \cite{Valentine16} showed 
that presence of even 1~\% of Sm vacancies
leads to smearing of the bulk hybridization gap in this material 
(being in qualitative agreement with earlier theoretical predictions \cite{Riseborough03}),
and would result in a breakdown of the topological Kondo insulator state \cite{Valentine16}.
Other studies of vacant samples Sm$_{1-x}$B$_6$, for $x$ ranging from $0.03$ to $0.2$, 
revealed resistivity saturation at lowest temperatures \cite{Gabani16} similar to that of the stoichiometric SmB$_6$. 
Because concentration of Sm vacancies in these samples substantially exceeds level of 1~\%, 
topologically protected surface states should not be present here, 
what implicates that
scenario of topological Kondo insulator is not applicable
in the case of all Sm-vacant samples.

The second above mentioned approach is represented by the scenario 
of valence-fluctuation induced hopping transport \cite{BaBa14} 
that is applicable also to vacant SmB$_6$-based samples. 
The key assumptions of this scenario, namely  
	(i) presence of valence fluctuations (VFs), 
	(ii) existence of an impurity band (IB),
and (iii) the Fermi level lying in the IB, 
 are in agreement with  theoretical studies of impurity states 
in mixed-valent materials \cite{Schlottman92,Schlottman96,Riseborough03},  
studies 
of electrodynamic properties of SmB$_6$ in the low-frequency regime 
 \cite{Gorshunov99},
optical conductivity studies of SmB$_6$ \cite{Laurita16,Laurita18}, 
observation of presence of variable-range hopping transport in SmB$_6$ at lowest temperatures \cite{Batkova18PhB}, 
and many other experimental observations reported for SmB$_6$
(see e.g.  \cite{Gorshunov99,Laurita16,Laurita18,Demishev18} and references therein).
As explained in \cite{BaBa14}, the energy of an impurity
in the semiconductor with an IB that contains metallic ions 
in two different valency states (e.g. $Me^{2+}$ and $Me^{3+}$) 
 depends on the distribution of
ion valences in the vicinity of this impurity
(because of different interaction energy of this impurity with ions 
of different valency
due to their different charge, ionic radii, and magnetic moment). 
Changes in distribution of ion valences 
unconditionally cause corresponding changes of 
the  impurity energies
and can be modeled by a rearrangement process (RP)
with a characteristic time constant, $t_{r}$ \cite{BaBa14}.
Thus, the energy $E_i$ of the impurity $i$ 
is not constant in time, but varies  within some interval 	
$\left\langle E_{i, min},E_{i, max} \right\rangle$ \cite{BaBa14}.
%
 Therefore, if the Fermi level, $E_F$, lies in the IB,
there has to exist a subnetwork of impurities for which it is true that due to
the RP
some occupied impurity energy levels can shift  
from the region below $E_F$ to the region above $E_F$, and analogously,
some empty levels from the region above $E_F$ can shift under $E_F$.
This not only creates favorable conditions for
temperature non-activated hops that are
responsible for temperature non-activated transport \cite{BaBa14}, but also causes  
a {\em change} of energy distribution function (EDF) of charge carriers. It is the purpose of this work to point out that the EDF of localized charge carriers in the ground state of valence fluctuating  semiconductors {\em qualitatively} differs from one expected for conventional semiconductors at $T=0$~K, 
and that the absence of low temperature resistivity divergence in (real) SmB$_6$-based
samples containing lattice imperfections can be a natural consequence of the changed EDF.

According to the ``quantum-limit'' hopping formula \cite{Ambegaokar71},
the intrinsic transition rate $\gamma_{ij}$ for an electron hopping from a site 
		$i$ with energy $E_i$ to an empty site $j$ with energy $E_j$ in the simplest case, 
when $|E_j-E_i|$ is of the order of the Debye energy or smaller,
and $kT$ is small compared to $|E_j-E_i|$,
		can be expressed as 
\begin{eqnarray}
		\gamma_{ij} =	\gamma_{0}e^{-2\alpha R_{ij} - (E_j-E_i)/kT} \mbox{ for } E_j > E_i\\
		\gamma_{ij} = \gamma_{0}e^{-2\alpha R_{ij}} \mbox{ for } E_j < E_i ,
				\label{TNAH}
	\end{eqnarray} 
		where $k$ is Boltzmann constant, $R_{ij}$ is the distance between sites $i$ and $j$, 
		and $\gamma_{0}$ is a constant, which depends 
		on the electron-phonon coupling strength, the phonon density of states,
		and other properties of the material, but which depends only weakly
		on the energies $E_i$ and $E_j$ or on $R_{ij}$  \cite{Ambegaokar71}. 
According to Eq.~(2), the intrinsic transition rate of  electron hop 
to a site of less energy decreases exponentially with increasing $R_{ij}$.
However, because $\gamma_{0}$ is finite,  $\gamma_{ij}$ must be also finite,
so there is always {\em non-zero} time interval $t_h$ until an electron can hop (tunnel)
to some empty site of less energy, while this time interval increases   
with increasing distance between the sites.
If the RP (e.g. represented by VFs) is the ground state property of a material,  
        finite $\gamma_{ij}$ suggests a non-zero probability 
				of finding some occupied energy levels above $E_F$ also 
				in the ground state.
In a conventional semiconductor (i.e. one without the RP)  
the time averaged occupation number of site $i$ 
in the thermal equilibrium (at neglecting electron-electron interactions
except those causing that not more than one electron can occupy a single site)
can be expressed in the form $\left\langle n_i\right\rangle = 1/\left\{1 +\exp [(E_i - E_F)/kT]\right\} $ 
\cite{Ambegaokar71},  
what for $T = 0$~K unconditionally means  
that all energy levels below $E_F$ are occupied  
and all  energy levels above $E_F$ are empty. 
Thus, the above mentioned non-zero probability to find occupied energy levels above $E_F$ 
in the ground state 
  implies  that  the ground-state EDF of the  subsystem of localized charge carriers 
	in materials with the RP 
		is {\em qualitatively} different from  the EDF in conventional semiconductors at $T=0$~K.
		This fundamental conclusion we support  by the following discussion.  

\begin{figure}[!h ]
			\center{
				\resizebox{1.00\columnwidth}{!}{%
  				\includegraphics{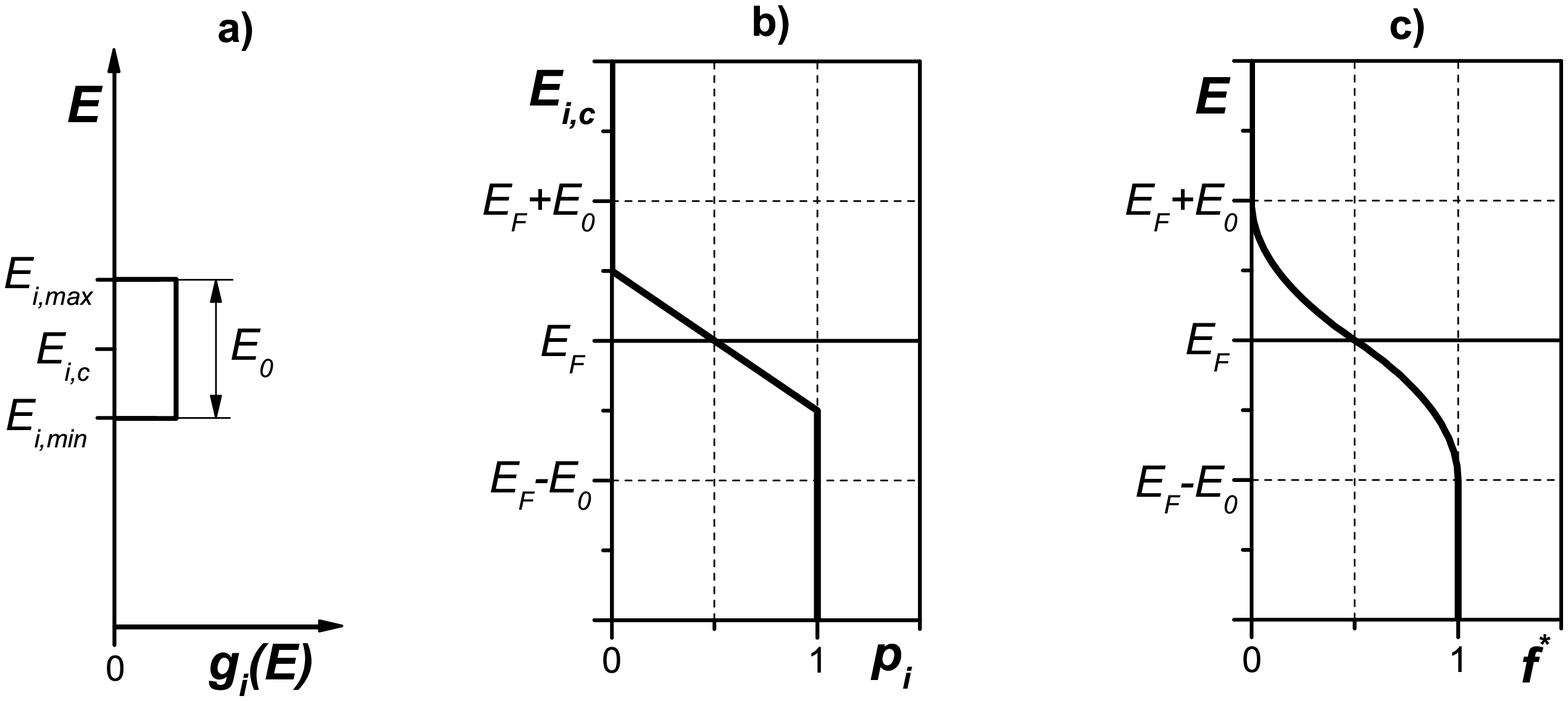}
        }
    	}
 			\caption
			{Schematic depiction of the partial DOS, $g_{i}(E)$, of the
			impurity center in the IB as introduced in \cite{BaBa14} (a), time averaged probability $p_{i}$ of the occupation the site $i$ having its energy interval centered at $E_{i,c}$ (b),
			and a time averaged probability of occupation of the state 
at energy $E$ in the IB (c).
       }
		\label{fig2new}
		\end{figure}

According to the scenario of valence-fluctuation induced hopping transport \cite{BaBa14},  
a hopping site $i$  can be characterized 
 by an energy interval of the typical width $E_{0} \approx E_{i,max} - E_{i,min}$
 and by the partial DOS, $g_{i}(E)$,  
			which is non-zero and constant within the 
				 interval $\left\langle E_{i, min},E_{i, max} \right\rangle$
				and zero outside it
				(see Fig.~\ref{fig2new}a).
Let us moreover characterize the site $i$  
				by a time averaged probability of the occupation of  
				this site by an electron, $ p_{i}\in \left\langle 0,1 \right\rangle $.
Because time interval $t_h$  until the electron can hop (tunnel)
to an empty site of less energy increases 
with increasing distance between the sites, 
at very low concentration of impurities, 
or in case of the RP with a very short $t_r$  (e.g. fast valence fluctuating process), 
it  can be reasonably considered that $t_r << t_h$.  
This limit case practically means that electron occupies the state for sufficiently long time
to receive any energy $E_{i}$ from the interval $\left\langle E_{i, min},E_{i, max} \right\rangle$. 	
We define the time averaged probability of the occupation of site $i$ 
				by an electron, $ p_{i} $, 	as the average value of the
				Fermi-Dirac distribution function (FDDF),
				$f_{0}(E_i,T) = 1/\left\{1 +\exp [(E_i - E_F)/kT]\right\}$, 
				over the energy interval $\left\langle E_{i, min},E_{i, max} \right\rangle$.
Assuming that all energy intervals have the typical width $E_0$ 
(i.e. $E_0 \equiv E_{i, max} - E_{i, min}$),
$p_{i}$ can be expressed in the form 
\begin{equation}
p_{i}(E_{i,c},T,E_0) = \frac{1}{E_{0}}\int_{E_{i,c}-E_0/2}^{E_{i,c}+E_0/2}f_{0}(E_i,T)dE_i,
\label{pi}
\end{equation}
where $E_{i,c} = (E_{i, min} + E_{i, max})/2$.
Fig.~\ref{fig2new}b shows $p_{i}$ as defined by Eq.~(\ref{pi})  
calculated
for $T=0$~K and for fixed parameter $E_0$.  
%
	As can be seen,
	$p_{i} = 1$ for  $E_{i,c} \leq E_F - E_{0}/2$, 
	$p_{i} = 0$ for  $E_{i,c} \geq E_F + E_{0}/2$, 
 and $p_{i}$ decreases linearly from 1 to 0 for $E_{i,c}$ 
	in the interval $E_F - E_{0}/2 < E_{c, i} < E_F + E_{0}/2$.
%
According to the above mentioned, all impurity sites having $E_{i,c}$ in the energy interval
$E-E_{0}/2 < E < E+E_{0}/2$ are overlapped at energy $E$,
and contribute to the probability of occupation of the state 
at energy $E$ proportionally to $p_{i}(E_{i,c},T,E_0)/E_0$.        
Thus we can determine the time averaged probability of 
occupation of the state at energy $E$ in the IB (i.e. the EDF) as
	\begin{equation}
		f^{*}(E,T,E_0) = \int_{E-E_{0}/2}^{E+E_{0}/2}\frac{p_{i}(E_{i,c},T,E_0)}{E_0}dE_{i,c} .
		\label{f0}
	\end{equation}
Analytical calculation of 
Eq.~\ref{f0} for $T=0$~K and for fixed $E_0$ gives 
\begin{equation}
f^{*} = 1 \mbox{ ~~for~~ } E \leq E_{F} - E_{0} \mbox{,} 
\end{equation}
\begin{equation}
f^{*} = \frac{-E^2 - 2E_0 E +E_0^2}{2E_0^2} \mbox{ for } E_{F} - E_{0} \leq E \leq E_{F} \mbox{,} 
\end{equation}
\begin{equation}
f^{*} = \frac{E^2 - 2E_0 E +E_0^2}{2E_0^2} \mbox{ for } E_{F} \leq E \leq E_{F} + E_{0} \mbox{,} 
\end{equation}
\begin{equation}
f^{*} = 0 \mbox{ ~~for~~ } E_{F} + E_{0} \leq E \mbox{,}
\end{equation}
as depicted in Fig.~\ref{fig2new}c. 
 As follows from  Eqs. (5-8), the $f^{*}(E, 0$~K$, E_0)$ is a continuous function
	with the finite slope $\partial f^{*}(E, 0$~K$, E_0)/\partial E = -1/E_0$ at the Fermi level.
This represents a qualitaive difference in comparison with FDDF, since  $\partial f_{0}(E, T)/\partial E = -1/4kT$ at $E_F$ (giving infinite slope for $T=0$~K). 
Considering that the slope of the EDF at $E_F$ reflects energy broadening 
(either due to a non-zero temperature or due to the RP), both slopes can be compared yielding 
\begin{equation}
	E_0 \approx 4kT.
\end{equation}

%
\begin{figure}[!t]
			\center{
				\resizebox{1.00\columnwidth}{!}{%
  				\includegraphics{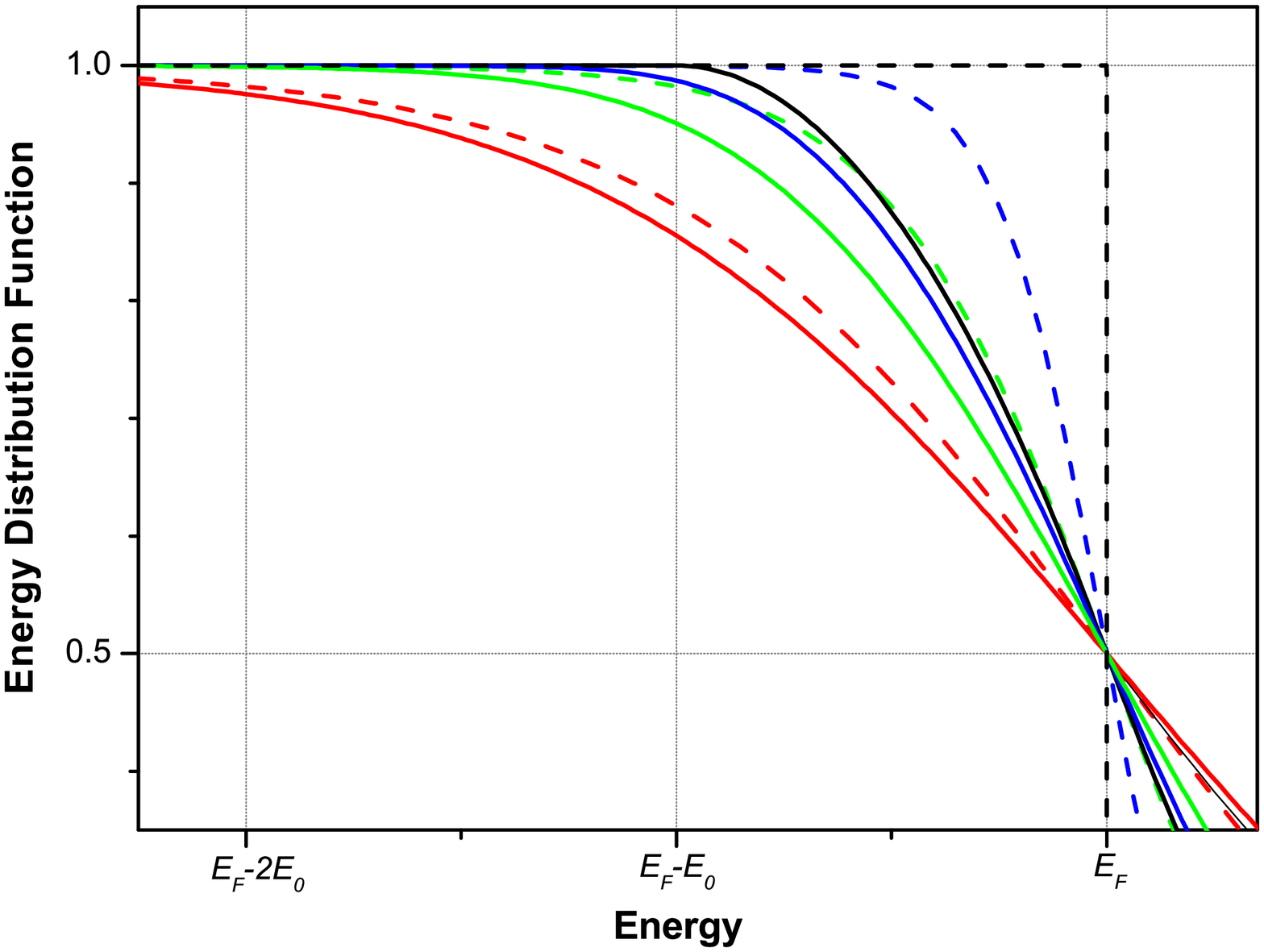}
        }
    	}
 			\caption{ Plots of $f^{*}$ (solid lines) compared with FDDFs (dashed lines) 
			for temperatures $T=0$~K (black), $T=E_0/8k$ (blue), $T=E_0/4k$ (green) and $T=E_0/2k$ (red).
}
		\label{FigNonFD}
		\end{figure}

Fig.~\ref{FigNonFD} shows numerically calculated plots of $f^{*}(E, T, E_0)$ 
for $T=0$~K, $T=E_0/8k$, $T=E_0/4k$ and $T=E_0/2k$
compared with FDDFs calculated for the same temperatures.
As can be seen, $f^{*}$ for $T=0$~K strongly differs from the FDDF
at  $T=0$~K, and can be much better approximated by the FDDF at the finite temperature $T=E_0/4k$.
As can be deduced from the plots for $T=E_0/8k$,  
the EDF of the system with the RP can  be
at sufficiently low temperatures
 much better approximated
by the FDDF at $T=E_0/4k$ than by the FDDF at real temperature of the system.
Thus it seems that 
the RP at lowest temperatures 
affects the EDF analogously as  temperature enhanced to the value of $E_0/4k$.
In general, all $f^{*}$ curves have less (negative) slope at $E_F$ in comparison with 
the slope of the FDDF (at $E_F$) at the same temperature,
resembling the
effect of a temperature enhancement due to the RP.

We suppose that the energy broadening as discussed above (for $t_r << t_h$) 
can be adopted also to the case of SmB$_6$ because of the 
fast valence fluctuation rates in this material 
\cite{Mock86}.
Deducing from Eqs. (5-8) and Fig.~\ref{FigNonFD}, at sufficiently high temperatures (where 
 $E_0$ is much less than thermal broadening) the EDF of localized charge carriers in SmB$_6$ will be practically unaffected by valence fluctuations (VFs). 
 Approaching temperatures $T\approx E_0/4k$ 
 the EDF will exhibit essential deviation from 
the FDDF due to  VFs, 
while at temperatures $T << E_0/4k$ the EDF will 
{\em resemble} the FDDF corresponding to significantly greater temperature, $T\approx E_0/4k$.    
Consequently, physical characteristics of SmB$_6$ that are governed by the EDF
(e.g. electrical resistivity) will for $T \to 0$~K converge to those expected at a non-zero temperature, providing another reason for the absence of the resistivity divergence. 

Here we would like to pay attention to some
possible new 
phenomena which could take place at 
temperatures around and below $T \approx E_0/4k$.
As follows from the definition (Eq.~\ref{f0}), $f^{*}(E,T,E_0)$  
is the function of parameter $E_0$, thus $E_0$ affects
the energy of the subsystem of localized charge carriers.
At the same time, $E_0$  reflects a "magnitude" 
of impurity energy fluctuation due to fluctuation 
of the properties of the surrounding lattice (because of VFs).
Thus, $E_0$ can represent a "driving force"
to minimize the total energy of the system.
For instance, "tuning" of the processes associated with VFs
(e.g. reasonable change of fluctuations in magnetic subsystem and the associated magnetic interactions/correlations) 
can yield such value of $E_0$ that the total energy 
will be minimized.
Although at high temperatures ($T>> E_0/4k$) $E_0$ will practically 
not affect the $f^{*}(E,T, E_0)$,
the influence of $E_0$ can become essential approaching temperatures close to $T \approx E_0/4k$, 
and it may appear as favorable to change physical state of the system such way that 
the total energy will be minimized due to changed interactions 
and correspondingly changed value of $E_0$.
No clear conclusion can be done here
whether eventual change of $E_0$ should be continuous in a certain temperature interval
or "sudden", resembling a phase transition. 
Nevertheless in can be still expected that such change of $E_0$ will be associated with 
change of physical properties of the system 
in a certain temperature region.  
As possible (or perhaps accidental) analogy with the above provided picture
we would like to mention very recent ESR studies of SmB$_6$ \cite{Demishev18}. 
The ESR signal observed at $T=5$~K was not more observed at $T=6$~K \cite{Demishev18};
moreover,
the critical behavior of the integrated ESR intensity,
$I(T) \sim (T^* -T)^\nu$, with characteristic temperature 
$T^*=5.34 \pm 0.05$~K and exponent $\nu=0.38\pm 0.03$ was observed,  
and the authors \cite{Demishev18} mentioned a possibility of 
"some abrupt structural/magnetic transition" to explain experimental results and emphasized  
the importance of re-considering "5~K anomaly" in context of modern topological physics
of SmB$_6$.

Finally, we would like to 
emphasize that presented results enable to explain
absence of resistivity divergence at lowest temperatures 
in both stoichiometric, as well as in vacant SmB$_6$ samples.
The scenario of valence-fluctuation induced hopping transport \cite{BaBa14} and its consequences discussed here 
shed new light onto
observations associated with "5~K anomaly" by inferring possible existence of new phenomena
"driven" by the subsystem of localized charge carriers,
while all these phenomena 
can be considered simultaneously with possible presence of topologically protected 
or/and polarity-driven metallic surface states in SmB$_6$.
We believe that our findings not only represent a base 
for  understanding the underlying physics  in valence fluctuating	 semiconducting  compounds
 at lowest temperatures, 
		but that they also indicate a necessity to consider similar phenomena in many other materials 
		with ''dynamical ground state'', 
		especially those obeying physical properties which cannot be adequately understood 
		presumably supposing the ground state being associated with the FDDF for absolute zero.   
%


This work was supported by the Slovak Scientific Agency VEGA (Grant No.~2/0015/17)
and by the Slovak Research and Development Agency (APVV-15-0115). 


\begin{thebibliography}{10}

\bibitem{Wachter93}
P.~Wachter.
\newblock {\em Handbook on the Physics and Chemistry of Rare Earths},
  volume~19.
\newblock North Holland, 1993.

\bibitem{Riseborough00}
Peter~S. Riseborough.
\newblock Heavy fermion semiconductors.
\newblock {\em Advances in Physics}, 49:257, 2000.

\bibitem{Allen79}
J.~W. Allen, B.~Batlogg, and P.~Wachter.
\newblock Large low-temperature {Hall} effect and resistivity in mixed-valent
  {SmB}$_{6}$.
\newblock {\em Phys. Rev. B}, 20:4807--4813, Dec 1979.

\bibitem{Cooley95}
J.~C. Cooley, M.~C. Aronson, Z.~Fisk, and P.~C. Canfield.
\newblock {SmB}$_{6}$: {Kondo} insulator or exotic metal?
\newblock {\em Phys. Rev. Lett.}, 74:1629 -- 1632, 1995.

\bibitem{Batko93}
I.~Bat$\!$'ko, P.~Farka\v{s}ovsk\'y, K.~Flachbart, E.~S. Konovalova, and Yu.~B.
  Paderno.
\newblock Low temperature resistivity of valence fluctuation compound
  {SmB}$_{6}$.
\newblock {\em Solid State Commun.}, 88:405 -- 410, 1993.

\bibitem{Gorshunov99}
B.~Gorshunov, N.~Sluchanko, A.~Volkov, M.~Dressel, G.~Knebel, A.~Loidl, and
  S.~Kunii.
\newblock Low-energy electrodynamics of {SmB}$_{6}$.
\newblock {\em Phys. Rev. B}, 59:1808--1814, Jan 1999.

\bibitem{Batkova18PhB}
M.~Batkova, I.~Batko, S.~Gab\'ani, E.~Ga\v zo, E.~Konovalova, and V.~Filippov.
\newblock Low temperature resistivity studies of {SmB}$_{6}$: Observation of
  two-dimensional variable-range hopping conductivity.
\newblock {\em Physica B}, 536:200--202, 2018.

\bibitem{Stern17}
A.~Stern, M.~Dzero, V.~M. Galitski, Z.~Fisk, and J.~Xia.
\newblock Surface-dominated conduction up to 240 k in the {Kondo} insulator
  {SmB}$_{6}$ under strain.
\newblock {\em Nature Materials}, 16:708 -- 711, 2017.

\bibitem{Batkova18Jal}
M.~Batkova, I.~Batko, F.~Stobiecki, B.~Szyma\'{n}skiand~P. Ku\'{s}wik,
  A.~Mackov\'a, and P.~Malinsk\'y.
\newblock Electrical properties of {SmB}$_{6}$ thin films \\prepared by pulsed
  laser deposition from a stoichiometric {SmB}$_{6}$ target.
\newblock {\em J. Alloys Compd.}, 744:821--827, 2018.

\bibitem{Batkova06}
M.~Batkova, I.~Batko, E.~S. Konovalova, N.~Shitsevalova, and Y.~Paderno.
\newblock Gap properties of {SmB}$_{6}$ and {YbB}$_{12}$: Electrical
  resistivity and tunnelling spectroscopy studies.
\newblock {\em Physica B}, 378 - 380:618, 2006.

\bibitem{Ioffe60}
A.F. Ioffe and A.R. Regel.
\newblock Non-crystalline, amorphous, and liquid electronic semiconductors.
\newblock {\em Progress in Semiconductors}, page 237, 1960.

\bibitem{Mott71}
N.F. Mott and E.~Davis.
\newblock {\em Electronic Processes in Non-Crystalline Materials}.
\newblock Clarendon Press, 1971.

\bibitem{Dzero10}
Maxim Dzero, Kai Sun, Victor Galitski, and Piers Coleman.
\newblock Topological {Kondo} insulators.
\newblock {\em Phys. Rev. Lett.}, 104:106408, Mar 2010.

\bibitem{Dzero12}
Maxim Dzero, Kai Sun, Piers Coleman, and Victor Galitski.
\newblock Theory of topological {Kondo} insulators.
\newblock {\em Phys. Rev. B}, 85:045130, Jan 2012.

\bibitem{Lu13}
Feng Lu, JianZhou Zhao, Hongming Weng, Zhong Fang, and Xi~Dai.
\newblock Correlated topological insulators with mixed valence.
\newblock {\em Phys. Rev. Lett.}, 110:096401, Feb 2013.

\bibitem{Alexandrov13}
Victor Alexandrov, Maxim Dzero, and Piers Coleman.
\newblock Cubic topological {Kondo} insulators.
\newblock {\em Phys. Rev. Lett.}, 111:226403, Nov 2013.

\bibitem{Zhu13}
Z.-H. Zhu, A.~Nicolaou, G.~Levy, N.~P. Butch, P.~Syers, X.~F. Wang,
  J.~Paglione, G.~A. Sawatzky, I.~S. Elfimov, and A.~Damascelli.
\newblock Polarity-driven surface metallicity in {SmB}$_{6}$.
\newblock {\em Phys. Rev. Lett.}, 111:216402, Nov 2013.

\bibitem{BaBa14}
I.~Batko and M.~Batkova.
\newblock {SmB}$_{6}$: Topological insulator or semiconductor with
  valence-fluctuation induced hopping transport?
\newblock {\em Solid State Commun.}, 196:18, 2014.

\bibitem{Wolgast13}
Steven Wolgast, Cagylian Kurdak, Kai Sun, J.~W. Allen, Dae-Jeong Kim, and
  Zachary Fisk.
\newblock Low-temperature surface conduction in the {Kondo} insulator
  {SmB}$_{6}$.
\newblock {\em Phys. Rev. B}, 88:180405, Nov 2013.

\bibitem{Kim13}
D.~J. Kim, S.~Thomas, T.~Grant, J.~Botimer, Z.~Fisk, and Jing Xia.
\newblock Surface {Hall} effect and nonlocal transport in {SmB}$_{6}$: Evidence
  for surface conduction.
\newblock {\em Scientific Reports}, 3:1, 2013.

\bibitem{Syers15}
Paul Syers, Dohun Kim, Michael~S. Fuhrer, and Johnpierre Paglione.
\newblock Tuning bulk and surface conduction in the proposed topological
  {Kondo} insulator {SmB}$_{6}$.
\newblock {\em Phys. Rev. Lett.}, 114:096601, Mar 2015.

\bibitem{Valentine16}
Michael~E. Valentine, Seyed Koohpayeh, W.~Adam Phelan, Tyrel~M. McQueen,
  Priscila F.~S. Rosa, Zachary Fisk, and Natalia Drichko.
\newblock Breakdown of the kondo insulating state in ${\mathrm{smb}}_{6}$ by
  introducing sm vacancies.
\newblock {\em Phys. Rev. B}, 94:075102, Aug 2016.

\bibitem{Riseborough03}
Peter~S. Riseborough.
\newblock Collapse of the coherence gap in {Kondo} semiconductors.
\newblock {\em Phys. Rev. B}, 68:235213, 2003.

\bibitem{Gabani16}
S.~Gab\'ani, M.~Orend\'a\v{c}, G.~Prist\'a\v{s}, E.~Ga\v{z}o, P.~Diko,
  S.~Piovar\v{c}i, V.~Glushkov, N.~Sluchanko, A.~Levchenko, N.~Shitsevalova,
  and K.~Flachbart.
\newblock Transport properties of variously doped {SmB}$_{6}$.
\newblock {\em Philosophical Magazine}, 96(31):3274--3283, 2016.

\bibitem{Schlottman92}
P.~Schlottmann.
\newblock Impurity bands in {Kondo} insulators.
\newblock {\em Phys. Rev. B}, 46:998, 1992.

\bibitem{Schlottman96}
P.~Schlottmann.
\newblock Influence of a {Kondo}-hole impurity band on magnetic instabilities
  in {Kondo} insulators.
\newblock {\em Phys. Rev. B}, 54:12324, 1996.

\bibitem{Laurita16}
N.J. Laurita, C.M. Morris, S.M. Koohpayeh, P.F.S. Rosa, W.A. Phelan, Z.~Fisk,
  T.M. McQueen, and N.P. Armitage.
\newblock Anomalous three-dimensional bulk ac conduction within the {Kondo} gap
  of {SmB}$_{6}$ single crystals.
\newblock {\em Phys. Rev. B}, 94:165154, 2016.

\bibitem{Laurita18}
N.J. Laurita, C.M. Morris, S.M. Koohpayeh, W.A. Phelan, T.M. McQueen, and N.P.
  Armitage.
\newblock Impurities or a neutral {Fermi} surface? {A} further examination of
  the low-energy ac optical conductivity of {SmB}$_{6}$.
\newblock {\em Physica B}, 536:78--84, 2018.

\bibitem{Demishev18}
S.~V. Demishev, M.~I. Gilmanov, A.~N. Samarin, A.~V. Semeno, N.~E. Sluchanko,
  N.~A. Samarin, A.~V. Bogach, N.~Yu. Shitsevalova, V.~B. Filipov, M.~S.
  Karasev, and V.~V. Glushkov.
\newblock Magnetic resonance probing of ground state in the mixed valence
  correlated topological insulator {SmB}$_{6}$.
\newblock {\em Scientific Reports}, 8:7125, 2018.

\bibitem{Ambegaokar71}
Vinay Ambegaokar, B.~I. Halperin, and J.~S. Langer.
\newblock Hopping conductivity in disordered systems.
\newblock {\em Phys. Rev. B}, 4:2612--2620, Oct 1971.

\bibitem{Mock86}
R.~Mock, E.~Zirngiebl, B.~Hillebrands, G.~G\"untherodt, and F.~Holtzberg.
\newblock Experimental identification of charge relaxation rates in
  intermediate-valence compounds by phonon spectroscopy.
\newblock {\em Phys. Rev. Lett.}, 57:1040--1043, Aug 1986.

\end{thebibliography}
\end{document}